\documentclass[twoside]{article}

\usepackage{graphics,graphicx,wrapfig,times}
\usepackage{color}
\usepackage{subeqn}

\newcommand{\be}{\begin{equation}}
\newcommand{\ee}{\end{equation}}
\newcommand{\la}{\langle}
\newcommand{\ra}{\rangle}
\newcommand{\bx}{{\bf{x}}}
\newcommand{\bX}{{\bf{X}}}
\newcommand{\obx}{\overline{\bf{x}}}
\newcommand{\ox}{\overline{x}}
\newcommand{\bn}{\widehat{\bf{n}}}
\newcommand{\bV}{{\bf{V}}}
\newcommand{\mD}{{\mathcal{D}}}
\newcommand{\mV}{{\mathcal{V}}}
\newcommand{\mL}{{\mathcal{L}}}
\newcommand{\mS}{{\mathcal{S}}}
\newcommand{\Erfc}{{\rm{Erfc}}}

\newcommand{\rmm}{{\rm{m}}}
\newcommand{\rms}{{\rm{s}}}

\usepackage[sc]{mathpazo} 
\usepackage[T1]{fontenc} 
\usepackage{microtype} 

\usepackage[hmarginratio=1:1,top=32mm,columnsep=20pt]{geometry} 
\usepackage[hang, small,labelfont=bf,up,textfont=it,up]{caption} 
\usepackage{booktabs} 
\usepackage{float} 
\usepackage{hyperref} 

\usepackage{lettrine} 
\usepackage{paralist} 

\usepackage{abstract} 

\usepackage{titlesec} 
\titleformat{\section}[block]{\large\scshape\centering}{\thesection.}{1em}{} 
\titleformat{\subsection}[block]{\large}{\thesubsection.}{1em}{} 

\usepackage{fancyhdr} 
\title{\vspace{-15mm}\fontsize{24pt}{10pt}\selectfont\textbf{
\vspace{-1.5truecm}
\\
{\small{CRS4 Technical Report 2012/PM12a, July 2012. Revised Version August 2014.}}
{\small{http://publications.crs4.it/pubdocs/2012/PM12a/pagnini\_massidda-levelset.pdf}}
\\
\vspace{1.0truecm}
Modelling turbulence effects\\ in wildland fire propagation\\
by the randomized level-set method\thanks{
This work has been started during a research period of Gianni Pagnini at CRS4   
supported by the Sardinian Regional Authority (PO Sardegna FSE 2007--2013, L.R. 7/2007).
} 
}}

\author{
\large
\textsc{Gianni Pagnini$^{1,2}$ and Luca Massidda$^3$}\\ [2mm]
\normalsize $^1$ BCAM - Basque Center for Applied Mathematics,\\
\normalsize Alameda de Mazarredo 14, 48009 Bilbao, Basque Country -- Spain\\
[1mm]
\normalsize $^2$
Ikerbasque, Basque Foundation for Science,\\
\normalsize Alameda Urquijo 36-5, Plaza Bizkaia, 48011 Bilbao, Basque Country -- Spain\\
[1mm]
\normalsize $^3$
CRS4 - Center for Advanced Studies, Research and Development in Sardinia,\\
\normalsize Polaris Ed. 1, 09010 Pula~(CA), Sardinia -- Italy\\ 
[1mm]
\normalsize {e-mail: gpagnini@bcamath.org; luca.massidda@crs4.it} 
\vspace{-5mm}
}

\date{}

\begin{document}

\maketitle 

\thispagestyle{fancy} 


\vspace{-0.1truecm}
\begin{abstract}
\noindent 
Turbulence is of paramount importance in wildland fire propagation since it randomly transports 
the hot air mass that can pre-heat and then ignite the area ahead the fire.
This contributes to give a random character to the firefront position
together with other phenomena as for example fire spotting, vegetation distribution (patchiness), 
gaseous combustion fluctuation, small-scale terrain elevation changes. Here only turbulence is considered.
The level-set method is used to numerically describe the evolution of the fireline contour 
that is assumed to have a random motion because of turbulence. 
The progression of the combustion process is then described by a level-set contour distributed according to a weight function given 
by the probability density function of the air particles in turbulent motion.
From the comparison between the ordinary and the randomized level-set methods, 
it emerges that the proposed modelling approach turns out to be suitable to simulate a moving firefront fed by the ground fuel and driven, 
beside the meteorological and orographical factors, also by the turbulent diffusion of the hot air. 
This approach allows the simulation of the fire overcoming of a firebreak zone. 
The discussed results are explorative and need to be subjected to a future validation.
\end{abstract}

\phantom{Key words.ABC}fire-atmosphere coupling, fire-induced flow, turbulence.

\vspace{0.1truecm}

\pagestyle{myheadings}
\thispagestyle{plain}
\markboth{G. Pagnini and L. Massidda}{
Modelling turbulence effects in wildland fire propagation
by the randomized level-set method}

\section{Introduction}
Modeling wildland fire propagation is a topic of interest for a number of reasons 
ranging from environmental motivations of wildland conservation and protection to the often unconsidered human safety and to property damage concern. 
Therefore, since several decades, the problem has been studied with many approaches 
depending on the theoretical knowledges and computational means of the times; 
the reader may refer for reviews to References 
\cite{pastor_etal-pecs-2003,perry-ppg-1998,sullivan-ijwf-2009a,sullivan-ijwf-2009b,sullivan-ijwf-2009c}. 

Wildland fire propagation is a complex multi-scale, as well as a multi-physics and multi-discipline process \cite{viegas-ptrsla-1998}, 
strongly influenced by the atmospheric wind. 
Since the firefront propagates at the ground level, 
apart from the fuel distribution, tipology and elevation as well as ground slope and orientation, 
it is influenced also by the dynamics of the Atmospheric Boundary Layer (ABL), whose flow is turbulent in nature.

Turbulence is of paramount importance in wildland fire propagation since it randomly transports 
the hot air mass that can pre-heat and then ignite the area ahead the fire.
This contributes to give a random character to the firefront position,
together with other phenomena such as fire spotting, vegetation distribution (patchiness), 
gaseous combustion fluctuation, small-scale ($10 \, \rmm$) terrain elevation changes just to name a few.

The ABL large-scale motion influences the firefront velocity by the mean wind while the turbulent small-scale motion plays an
important role in the wildfire spreading \cite{filippi_etal-james-2009,mandel_etal-gmd-2011,sun_etal-ijwf-2009}.
Actually, when a wildland fire occurs, the ABL is forced also by the fire-atmosphere coupling and, close to the firefront, 
by the fire-induced flow, so that turbulence intensity increases.
Accounting for the effects of turbulence on the fire propagation can improve operational models. 

The wildland fire is fed by the fuel on the ground and driven, beside meteorological and orographical factors,
also by the heat transfer that pre-heats the fuel and aids the fire propagation. 
The heat flux is turbulent, 
because of the turbulent ABL and of the turbulence generated by the fire-atmosphere coupling \cite{clements_etal-jgr-2008}.
Then, since the dependence of the fire propagation on the heat flux \cite{clark_etal-jam-1996}, 
the turbulent heat transport gives a random character to the firefront trajectory.
Spatial (horizontal and vertical) scales interested by turbulence are 
ranging from $1 \, \rmm$ to $10^3 \, \rmm$ \cite[Table 1]{sullivan-ijwf-2009a}.

Wildland fire propagation has been recently modelled using reaction-diffusion type equations, 
see e.g. \cite{babak_etal-siamjam-2009,mandel_etal-mcs-2008}, percolation theory, see e.g. \cite{favier-pla-2004,hunt-c-2007},
stochastic approaches, see e.g. 
\cite{almeida_etal-jpcs-2011,boychuk_etal-ees-2009,perryman_etal-ijwf-2013}, 
small-world network, see e.g. \cite{zekri_etal-pre-2005}, and the level-set method, 
see e.g. \cite{beezley_etal-lncs-2008,dobrinkova_etal-lncs-2011,mallet_etal-cma-2009,mandel_etal-ieee-2009,mandel_etal-gmd-2011}. 
Here, in order to physically model the global effects of turbulence on the firefront propagation, 
the suitability of a recent approach \cite{pagnini_etal-prl-2011} based on the statistical distribution of the level-set contour 
is investigated. 

The level-set is a powerful method to track moving interfaces \cite{sethian_etal-arfm-2003} that 
allows the representation of the burning region on a simple Cartesian grid and the flexible implementation of various ignition modes. 
Moreover, this method is particularly appropriate to handle problems that arise from propagation of wildfires 
because it leads to an accurate calculation of the front normal vector, which is necessary to compute the Rate Of Spread (ROS) of the fire. 
The level-set method can automatically deal with topological changes that can occur during the fire spreading, 
as the merging of separate flame fronts or the formation of unburned ``islands''. 
The motion of the level-set contour is here assumed to be random and distributed according 
to the probability density function (PDF) of the turbulent displacement of the hot air particles.
For this reason the present approach has been named {\it randomized level-set method}. 
Since statistically the particle PDF follows from an ensemble average, 
the resulting effective fireline contour follows as well from an ensemble average of random fireline contours.

In Section \ref{sec: model} 
an approach to include turbulence effects into the level-set method is introduced. 
A model that takes into account the pre-heating induced by the hot air turbulent flow is also proposed.
In Section \ref{sec: planefront} the literature formulation based on the ordinary level-set method and the present formulation
are compared with respect to the expansion of the burned area.
In Section \ref{sec: application} the numerical scheme for simulations is illustrated and numerical results are shown.
Simulated case studies are focused on the differences with the ordinary level-set approach and 
on the successfully tackle of a realistic situation as fire overcoming of firebreaks by heat convection.

\section{Model description}
\label{sec: model}
\subsection{The level-set method}
The level-set method was originally introduced by Osher \& Sethian \cite{osher_etal-jcp-1988}. 
This approach is particularly useful to handle problems in which the speed of the evolving interface is 
dependent on interface's properties such as curvature and normal direction, as well as the 
boundary conditions at the interface location.
Hence, it is suitable for problems in which the topology of the evolving interface 
changes during the process and for problems in which sharp corners and cusps can be 
generated.

Let $\Gamma(t)$ be the fireline contour then, in a two dimensional domain, 
it can be represented by an isoline of an auxiliary function $\gamma(\bx,t)$, 
such that $\bx \in \mS \subseteq R^2$ and $\gamma: \mS \times [0,+\infty[ \to R$, 
as follows $\Gamma(t) = \{\bx \in \mS | \gamma(\bx,t) = \gamma_0 = constant\}$. 
Then the evolution in time of the isoline is given by
\be
\frac{D \gamma}{D t}=\frac{\partial \gamma}{\partial t}+\frac{d \bx}{dt} \cdot \nabla 
\gamma = \frac{D \gamma_0}{D t}=0 \,.
\label{Eq: ls0}
\ee
If the motion of the surface points is directed along the normal direction then
\be
\frac{d \bx}{dt} = \bV(\bx,t)= \mV(\bx,t) \, \bn \,, \quad
\bn=-\frac{\nabla \gamma}{||\nabla \gamma||} \,,
\ee
and (\ref{Eq: ls0}) becomes the {\it ordinary} level-set equation
\be
\frac{\partial \gamma}{\partial t} = \mV(\bx,t) \, ||\nabla \gamma || \,.
\label{Eq: ls}
\ee

For our purpose we take $\mV(\bx,t)$ as the ROS of the firefront, i.e. 
the modulus of the velocity at which the fire contour propagates along its normal.
Let $\varphi(\bx,t)$ be an indicator function which takes values $0$ for unburned points 
and $1$ for burned points, then the area burned by the wildland fire may be defined as 
$\Omega(t)=\{\bx \in \mS | \varphi(\bx,t)=1 \}$ and it holds
\be
\varphi(\bx,t)=
\left\{
\begin{array}{lr}
1 \,, & \bx \in \Omega(t) \\
\\
0 \,, & \bx \not\in \Omega(t) \\
\end{array}
\right.
\,.
\label{PBeq00}
\ee
The boundary of $\Omega$, i.e. $\partial \Omega(t)$, is $\Gamma(t)$ that is the front line contour of the wildland fire.
When the ROS $\mV(\bx,t)$ is known,  
the evolution of the firefront can be efficiently simulated by the numerical solution of (\ref{Eq: ls}).

The ROS value essentially depends on wind intensity, 
on the orography of the terrain, on the type and conditions of the vegetation over which the fire is spread. 
Through experimental campaign and some physical modelling of the spread mechanism,
several formulae for the ROS have been derived in the recent past. 
The Rothermel expression \cite{rothermel-1972} has gained much attention and has been applied with success.
In our numerical simulation we will use Rothermel formula, but the present approach is valid for any formula of the ROS.

\subsection{The randomized level-set method}
A new approach for modelling wildland fire propagation 
based on the level-set method has been preliminarily proposed by the authors 
\cite{pagnini_etal-mfbr-2012}.
This approach, named the {\it randomized level-set method}, aims to include turbulence effects.
Effects due to the ABL turbulence and due to the fire-induced turbulent flow close to the flame front
are taken into account by means of a single global turbulent heat transfer.

The concept of the randomized level-set method is founded on the idea that, due to the 
turbulence, the firefront contour cannot be assumed to be deterministic.
Hence, the resulting firefront propagates randomly.
After ensemble averaging, a statistically distributed fireline follows that describes an {\it effective} fireline contour.

Let $\obx(t,\obx_0)$ be a deterministic trajectory with initial condition $\obx_0$, i.e. 
$\obx(0,\obx_0)=\obx_0$, and driven solely by the deterministic velocity field $\bV(\obx,t)$.
Moreover, let $\bX^\omega(t,\obx_0)=\obx(t,\obx_0)+\chi^\omega$ be the $\omega$-realization of a 
random trajectory driven by the noise $\chi$, with average value $\la 
\bX^\omega(t,\obx_0) \ra=\obx(t,\obx_0)$ and the same fixed initial condition 
$\bX^\omega(0,\obx_0)=\obx(0,\obx_0)=\obx_0$ in all realizations.
Hence, the $\omega$-realization of the fireline contour follows to be
\be
\varphi^\omega(\bx,t)=\int_{R^2} \varphi(\obx_0,0) \, \delta(\bx-\bX^\omega(t,\obx_0)) \, d 
\obx_0 \,.
\label{varphiomega0}
\ee
Since the trajectory $\obx(t,\obx_0)$ is time-reversible,
i.e. the Jacobian $J$ of the evolution from $\obx(0,\obx_0)=\obx_0$ to $\obx(t,\obx_0)$ is $J=\displaystyle{\frac{d \obx_0}{d \obx} \ne 1}$,
and setting an incompressibility-like condition, i.e. $J=1$,
formula (\ref{varphiomega0}) becomes
\be
\varphi^\omega(\bx,t)=\int_{R^2} \varphi(\obx,t) \, \delta(\bx-\bX^\omega(t,\obx)) \, d \obx \,.
\label{varphiomega}
\ee
Finally, after averaging, the effective firefront contour is determined as
\begin{eqnarray}
\la \varphi^\omega(\bx,t) \ra &=&
\la \int_{R^2} \varphi(\obx,t) \, \delta(\bx - \bX^\omega(t,\obx)) \, d \obx \ra \nonumber \\
&=& \int_{R^2} \varphi(\obx,t) \, \la \delta(\bx-\bX^\omega(t,\obx)) \ra \, d\obx \nonumber \\
&=& \int_{R^2} \varphi(\obx,t) \, p(\bx;t|\obx) \, d \obx \nonumber \\
&=& \int_{\Omega(t)} p(\bx;t|\obx) \, d\obx = \varphi_e(\bx,t) \,,
\label{PBeq}
\end{eqnarray}
where $p(\bx;t|\obx)=p(\bx-\obx;t)$ is the PDF of the turbulent dispersion of the hot flow particles
with average position $\obx$. Last equality but one follows from the definition of $\varphi(\obx,t)$, see (\ref{PBeq00}).
Formula (\ref{PBeq}) has been originally proposed to model the burned mass fraction in 
turbulent premixed combustion \cite{pagnini_etal-prl-2011}.

It is here remarked that the relation between the present model and that discussed in \cite{pagnini_etal-prl-2011} 
is purely at the level of the mathematical formalism.
No relationship is established between the present effective fire contour in wildland fire propagation
and the burned mass fraction in turbulent premixed combustion as discussed in \cite{pagnini_etal-prl-2011}.

It is worth-noting to remark that the deterministic trajectory $\obx$ is the trajectory of a point 
belonging to the ordinary level-set contour with the initial condition $\obx_0$.
Since the deterministic motion is recovered when $p(\bx-\obx;t)=\delta(\bx-\obx)$,
formula (\ref{PBeq}) gives
\be
\int_{\Omega(t)}\delta(\bx-\obx) \, d\obx = \varphi(\bx,t)=
\left\{
\begin{array}{lr}
1 \,, & \bx \in \Omega(t) \\
\\
0 \,, & \bx \not\in \Omega(t) \\
\end{array}
\right.
\,.
\label{PBeq0}
\ee
Finally, combining (\ref{PBeq}) and (\ref{PBeq0}) it follows that
\be
\varphi_e(\bx,t)=\int_{R^2} p(\bx;t|\obx) \varphi(\obx,t) \, d\obx \,.
\label{PBeq2}
\ee

By applying the Reynolds transport theorem to (\ref{PBeq}), the evolution equation of the 
effective firefront $\varphi_e(\bx,t)$ can be obtained as follows \cite{pagnini_etal-prl-2011}
\begin{eqnarray}
\frac{\partial \varphi_e}{\partial t}
&=& \frac{\partial}{\partial t} \int_{\Omega(t)} p(\bx;t|\obx) \, d\obx \nonumber \\
&=& \int_{\Omega(t)} \frac{\partial p}{\partial t} \, d\obx + 
\int_{\partial\Omega(t)} p(\bx;t|\obx) \, (\bV(\obx,t) \cdot \bn(\obx,t)) \, d{\bf{s}} \nonumber \\
&=& \int_{\Omega(t)} \frac{\partial p}{\partial t} 
\, d \obx
+ \int_{\Omega(t)} \nabla_{\obx} \cdot [\bV(\obx,t) \, p(\bx-\obx;t)] \, d \obx \,,
\label{reynolds0}
\end{eqnarray}
where $\partial\Omega(t)$ is the boundary of $\Omega(t)$ with infinitesimal element $d{\bf{s}}$, 
normal $\bn$ and velocity modulus determined by the ROS, i.e. $\bV(\bx,t)=\mV(\bx,t) \, \bn$.
From the second to the third line, the divergence theorem has been used in the second addendum of the RHS.
By introducing the mean front curvature $\kappa(\obx,t)=\nabla_{\obx} \cdot \bn/2$, 
equation (\ref{reynolds0}) becomes
\be
\frac{\partial \varphi_e}{\partial t} =
\int_{\Omega(t)} \frac{\partial p}{\partial t} \, d \obx
+ \int_{\Omega(t)} \bV \cdot \nabla_{\obx} \, p \ d\obx 
+ \int_{\Omega(t)} p \, \left\{
\frac{\partial \mV}{\partial \kappa} \nabla_{\obx} \, \kappa \cdot \bn + 2 \, 
\mV(\kappa,t) \, \kappa(\obx,t) \right\} \, d\obx \,.
\label{reynolds}
\ee

Fireline propagation follows to be driven by the turbulent dispersion (i.e. $p(\bx-\obx;t)$),
the velocity field (i.e. $\bV(\obx,t)$) and the mean front curvature (i.e. $\kappa(\obx,t)$).
Hereinafter, points $\bx$ such that $\varphi_e(\bx,t) > 0.5$ are marked as burned and 
the effective burned area follows to be $\Omega_e(t)=\{\bx \in \mS | \varphi_e(\bx,t) > 0.5\}$.
For a deterministic motion, i.e. $p(\bx-\obx;t)=\delta(\bx-\obx)$, 
equation (\ref{reynolds}) reduces to the ordinary level-set equation (\ref{Eq: ls}) \cite{pagnini_etal-prl-2011}.

\subsection{The heating-before-burning law}
The model is completed by introducing a law to describe the heat transferred to the surrounding fuel.
In the present approach, hot air is considered to be an heat source and to be transferred by turbulence
with proper spatial scales ranging from $1 \, \rmm$ to $10^3 \, \rmm$ \cite[Table 1]{sullivan-ijwf-2009a}.
Heat removes the moisture from the fuel enabling the fire to propagate more easily.
Roughly speaking, when the fuel is heated its temperature starts to increase according to its specific heat
until it reaches an ignition temperature and combustion can begin. 

This process may be simply viewed as an accumulation process that can be described by a function $\psi(\bx,t)$, 
with initial condition $\psi(\bx,0)=0$ corresponding to the initial unburned fuel.

Relating the accumulation function $\psi(\bx,t)$ with the amount of heat $\Delta Q$, 
since the increasing of the fuel temperature $T(\bx,t)$ is 
proportional to $\Delta Q$ through the heat capacity, it holds
\be
\psi(\bx,t) \propto \Delta Q \propto \frac{T(\bx,t)-T(\bx,0)}{T_{ign}-T(\bx,0)} \,,
\label{DeltaQ}
\ee
where $T_{ign}$ is the ignition temperature.

Moreover, let function $\psi(\bx,t)$
be assumed to be proportional to the accumulation in time of the effective firefront $\varphi_e(\bx,t)$, i.e.
\be
\int_0^t \frac{\varphi_e(\bx,\xi)}{\tau} \, d\xi = \psi(\bx,t) \,,
\label{restart0}
\ee
where $\tau$ is a {\it characteristic time} that embodies properties concerning increasing of fuel temperature as caused by convected heat.

Finally, let $\Delta t$ be the auto-ignition delay,
i.e. the elapsed time after which it is met the auto-ignition condition $T(\bx,\Delta t)=T_{ign}$,
then it can be stated that 
\be
{\rm if} \quad \psi(\bx,\Delta t) = \int_0^{\Delta t} \frac{\varphi_e(\bx,t)}{\tau} \, dt=1  \quad {\rm then} \quad \varphi(\bx,\Delta t) = 1 \,.
\label{restart}
\ee
In the present oversimplified framework where only turbulence is considered, 
auto-ignition condition is referred to spontaneous ignition without flames and sparks.
Please note that condition (\ref{restart}) is applied to $\varphi(\bx,t)$ that is the underlying burning indicator.
Moreover, from (\ref{DeltaQ}) and (\ref{restart}), 
the governing equation for the temperature field follows to be
\be
\frac{\partial T(\bx,t)}{\partial t} \propto \varphi_e(\bx,t) \, \frac{T_{ign}-T(\bx,0)}{\tau} \,, \quad
T \le T_{ign} \,.
\label{def: tau}
\ee

In this approach, two competing mechanism are therefore acting for the firefront propagation:
the direct advancement according to the ROS and  
the auto-ignition caused by the turbulent diffusion of hot air.

\section{Comparison between the ordinary and the present formulations}
\label{sec: planefront}
In order to discuss the comparison between the ordinary and the present formulations, 
let us introduce the following terminology:

\smallskip
\noindent
$-$ firefront propagation is labelled as 
``cold'' 
when the fireline contour is solely determined by the burning criterion $\varphi_e(\bx,t) > 0.5$, 
without considering the heating-before-burning law (\ref{def: tau}), 

\smallskip
\noindent
$-$ firefront propagation is labelled as
``hot'' when the burned-area growing is determined according to both the burning criteria: 
$\varphi_e(\bx,t) > 0.5$ and the heating-before-burning law (\ref{def: tau}).

Let us assume the following simple isotropic parabolic model for turbulent diffusion of the hot air mass around 
the average fireline $\obx$, i.e. 
\be
\frac{\partial p}{\partial t}=\mD \, \nabla^2 p \,, \quad p(\bx-\obx,0)=\delta(\bx-\obx) \,,
\label{parabolic}
\ee
where $\mD$ is the diffusion coefficient.
Solution of (\ref{parabolic}) is
\be
p(\bx-\obx;t)= \frac{1}{4 \pi \mD t} \exp\left\{-\frac{(x-\overline{x})^2+(y-\overline{y})^2}{4 \mD t} \right\} \,,
\label{gaussian2D}
\ee
and the particle displacement variance is related to the turbulent diffusion coefficient $\mD$ by
$\la (x-\overline{x})^2 \ra=\la (y-\overline{y})^2 \ra = 2 \mD t$.

When the normal to the front $\bn$ is constant the curvature $\kappa$ is null.
In this case 
the process reduces to a one-dimensional problem and solution of (\ref{reynolds}) is \cite{pagnini_etal-prl-2011}
\be
\varphi_e(x,t)=
\frac{1}{2}
\left\{\Erfc\left[\frac{x-\mL_R(t)}{2 \, \sqrt{\mD \, t}}\right]
- \Erfc\left[\frac{x-\mL_L(t)}{2 \, \sqrt{\mD \, t}}\right]
\right\} \,,
\label{c}
\ee
where $\Erfc$ is the complementary Error function, $\mL_R$ and $\mL_L$ are the right and left fronts, respectively, 
i.e. $\Omega(t)=[\mL_L(t);\mL_R(t)]$, which are defined by
\be
\frac{d\mL_R}{dt}=-\frac{d\mL_L}{dt}=\mV(t) \,.
\ee
Moreover, let $\mV=constant$, then the right-side firefront position determined by the ordinary level-set equation is 
$\ox(t)=\mL_R=\mL_{R0} + \mV \, t$.
 
Let us start considering the ``cold'' propagation.
In order to compare the propagation of the ``cold'' front with the ordinary level-set approach, the
effective front line is computed in $x=\mL_R$ and it holds
\be
\varphi_e(\mL_R,t)=
\frac{1}{2}
\left\{1
- \Erfc\left[\frac{\mL_{R0}-\mL_{L0} + 2 \, \mV \, t}{2 \, \sqrt{\mD \, t}}\right]
\right\} < \frac{1}{2} \,, \quad 0 < t < \infty \,.
\ee
Which it means that if the ``cold'' front marks a point as burned when $\varphi_e > 0.5$,
then the ``cold'' front propagation is always slower than the ordinary level-set front propagation.
This is due to the turbulent diffusive mechanism associated to the randomized approach.
In fact, from (\ref{PBeq}) and (\ref{PBeq0}) it follows that
\be
\int_{-\infty}^{+\infty} \varphi_e(x,t) \, dx = 
\int_{-\infty}^{+\infty} \varphi(x,t) \, dx = M(t) \,,
\ee
because $\displaystyle{\int_{-\infty}^{+\infty} p(x;t|\ox) \, dx=\int_{-\infty}^{+\infty} \delta(x-\ox) \, dx=1}$.
In the ordinary case, quantity $M(t)$ is fully located, at any instant $t$, in a finite domain $\Omega(t)$ 
while in the randomized case it is spread over the infinite domain $R^2$. This spreading generates the inequalities:
\begin{subequations}
\be
\varphi_e(x,t) < \varphi(x,t)=1 \,, \quad x \in [\mL_L(t); \mL_R(t)] \,, 
\ee
\be
\varphi_e(x,t) > \varphi(x,t)=0 \,, \quad x \not\in [\mL_R(t); \mL_R(t)] \,.
\ee
\end{subequations}

Let us consider now the ``hot'' propagation.
Assuming that with the ordinary level-set method a point $\bx$ is instantaneously burned when the firefront reaches it.
Then, each point turns to burned 
after an elapsed time $\delta t$ according to the kinematic law $\bx=\bx_0 + \int_0^{\delta t} \bV(\bx,\xi) \, d\xi$.
Since $\varphi=1$ for burned points,
from (\ref{restart}) the travelling time $\delta t$ corresponds also to the characteristic time $\tau$,
i.e. $\delta t=\tau$.
This means that the ``hot'' front propagation is faster than the ordinary level-set front when 
the auto-ignition delay $\Delta t$, computed according to the heating-before-burning law (\ref{restart}),
is less than the ordinary level-set travel-time $\delta t$, i.e. $\Delta t < \tau=\delta t$.
In fact, if (\ref{restart}) is met in a temporal interval $\Delta t < \tau=\delta t$, 
then the point under consideration is marked as burned by the present approach sooner than by the ordinary level-set approach.

In integral form, equation (\ref{parabolic}) reads
\be
p(\bx;t|\obx)= \delta(\bx-\obx) + \mD \int_0^t \nabla^2 p(\bx;\xi|\obx) \, d\xi \,.
\label{parabolic-integralform}
\ee
Then, by using (\ref{PBeq2}) and (\ref{parabolic-integralform}), 
the heating-before-burning criterion (\ref{restart}) for $\psi(\bx,\Delta t)=1$ can be re-written as
\begin{eqnarray}
\tau &=& \int_0^{\Delta t} \left\{\int_{R^2} \delta(\bx-\obx) \varphi(\obx,t) \, d\obx \right\} \, dt 
+ \mD \int_0^{\Delta t} \left\{\int_{R^2}\left[\int_0^t \nabla^2 p(\bx;\xi|\obx) \, d\xi \right]
\varphi(\obx,t) \, d\obx \right\} \, dt \,, \nonumber \\
\nonumber \\
&=& \Delta t + \mD \int_0^{\Delta t} \left\{\int_{R^2}\left[\int_0^t \nabla^2 p(\bx;\xi|\obx) d\xi \right] 
\varphi(\obx,t) \, d\obx \right\} \, dt \,, 
\end{eqnarray}
and it follows that the ``hot'' front is faster than the ordinary one,
i.e. the inequality $\Delta t < \tau$ holds, when the condition $\nabla^2 p > 0$ is met. 
Moreover larger is $\mD$ smaller is $\Delta t$, such that the difference $\tau-\Delta t$ increases. 
In the simple Gaussian model (\ref{gaussian2D}) here considered,   
condition $\nabla^2 p >0$ is met by those points that are located in $\bx > \obx + \sqrt{2 \mD \Delta t}$. 
So, in an elapsed time $\Delta t < \tau$, the ``hot'' front ignites solely such domain
and the time interval $\Delta t$ decreases when $\mD$ increases. 
Actually, the points in the domain defined by $\nabla^2 p \le 0$ turn to burned according to the propagation of the ``cold'' front.

Finally, stronger is the turbulence more distant from the fire flame the hot air is diffused.
Hence, the pre-heating action can ignite even very far away from the level-set fireline.
This fact is an acceleration factor for the firefront propagation.  

\section{Numerical simulations}
\label{sec: application}
\subsection{Numerical algorithm}
There are several numerical methods that can be and have been applied to simulate firefront propagation. 
These methods depend on available computational power and on purposes of the analysis.
The level-set method naturally appears as a good choice because it is an Eulerian method, 
it does not suffer from any topological issue, 
the line of the firefronts may freely merge or divide, 
it is accurate in the description of the front and of its normal, 
it is also straightforward in its implementation in a structured grid with finite difference numerical approximation.

We consider a rectangular two-dimensional domain on which  
we define a regular uniform Cartesian grid  
having $N_x \times N_y$ points $\bx_{ij}$ with origin ${\bf O} = (O_x, O_y)$ and grid spacing $h$, i.e. 
\begin{subequations}
\begin{eqnarray}
& & \bx_{ij} = (x_i, y_j) \,, \\ 
& & x_i = O_x + ih \quad {\rm with} \quad i=0 \ldots N_x -1 \,, \\
& & y_i = O_y + jh \quad {\rm with} \quad j=0 \ldots N_y -1 \,.
\end{eqnarray}
\end{subequations}
Each quantity that depends on space is defined on this set of points.

The finite difference approach allows to have a numerical approximation of the differential operators 
involved into the partial differential equations under consideration. Several numerical schemes are possible, 
each one with different accuracy, stability and computational effort. 
We have adopted an Essentially Non Oscillatory scheme of the first order for spatial derivatives 
to have some upwinding and a stable scheme.

A Total Variation Diminishing (TVD) Runge--Kutta scheme of the second order is adopted for time advancing, 
an uniform sampling of time is used. 
The second order integration in time was chosen in order to minimize the numerical diffusion of the algorithm 
that would affect the modelled diffusion process.
The scheme is explicit and subjected to the Courant--Friedrichs--Lewy (CFL) condition, that relates the maximum allowable time step $dt$
with the front speed $\mV(\bx,t)$ and the grid spacing $h$, i.e.
\be
dt < \frac{\min{h}}{\max{\mV(\bx, t)}} \,, \quad \forall \bx \,.
\label{dt}
\ee
Since the velocity of the firefront is in principle variable in time, the maximum allowable time step is calculated at each time step.

Given a non uniform initial value of the level-set function,
the steps of the numerical procedure are:

\smallskip

\begin{enumerate}
\item The central difference approximation of the gradient of the level-set function is calculated.
\item The gradient is normalized to obtain the unit normal to the front.
\item 
The ROS is calculated in each point of the Cartesian grid by
using data of the wind and the terrain conditions at each point and also the orientation of the firefront. 
\item By using an upwind approximation of the gradient of the level-set function,
the normal velocity term is added to the RHS of the equation. 
\item The first stage of the TVD Runge Kutta scheme is completed to obtain an approximation of the new value of the level-set function 
for the next time step.
\item Steps from 1 to 4 are repeated using the new value of the level-set function.
\item The second stage of the TVD Runge Kutta scheme gives the new value of the level-set function.
\item The new value of $\varphi_e$ is calculated through numerical integration 
of the product of $\varphi$ times the PDF of particle distribution as stated in (\ref{PBeq2}).
\item Function $\psi$ is updated for each point by integration in time with the current value of $\varphi_e$, see (\ref{restart}).
\item In any point with $\psi > 1$, the ignition is possible and the value of the level-set function $\varphi$ is updated to allow ignition.
\item Current time is updated as well as the level-set function. 
The new value of the maximum allowable time step is calculated through the CFL condition and the operations are repeated for a new time step.
\end{enumerate}

\smallskip

It is worth noting to remark that step $8$ of this numerical procedure, which corresponds to formulation stated in (\ref{PBeq2}), 
is strongly close to the Smoothed Particle Hydrodynamics (SPH) \cite{monaghan-rpp-2005}.
But, with respect to the SPH method, in the present approach the choice of the kernel function and the smoothing length are removed because 
they straightforwardly follow from the particle PDF. 
Actually, by assuming here a parabolic model for turbulent diffusion (\ref{parabolic}), in terms of SPH approach, 
the kernel function is the Gaussian PDF (\ref{gaussian2D}) and the smoothing length is equal to $4 \mD t$.

\subsection{Numerical simulation set-up}
Simulations are performed assuming parabolic model (\ref{parabolic})
for the turbulent diffusion.
All processes from large to small-scales of motion which are related to the turbulent heat transfer,
i.e. from the Atmospheric Boundary Layer to the fire-induced flow,                                                                 
are here represented in an oversimplified way by using a single turbulent diffusion coefficient $\mD$.
Quantitative estimation of turbulence generated by fire is still an open issue \cite{clements_etal-jgr-2008,seto_etal-afm-2013}.
Simulations are mainly intended to investigate the capacity of the proposed approach.                                       
In particular, since it is well-known that the value of thermal diffusity in ambient air is around $2 \times 10^{-5} \, \rmm^2\rms^{-1}$,
the effect of turbulence is here accounted for generating a constant turbulent diffusion coefficient 
of three and four orders of magnitude higher, i.e. 
$\mD = 4 \times 10^{-2} \, \rmm^2\rms^{-1}$, $1.5 \times 10^{-1} \, \rmm^2\rms^{-1}$, $3.5 \times 10^{-1} \, \rmm^2\rms^{-1}$.

Values of the characteristic time $\tau$ have been chosen with the only intention to select
case studies useful to highlight the potentiality of the proposed approach.
As a consequence of the considered parabolic model (\ref{parabolic}), 
the hot air is instantaneously spread over an infinite domain so that the whole fuel ground is heated by convection.
The characteristic time $\tau$ represents the inertia of the fuel to the auto-ignition caused by the convected heat.
In fact, the lower is the value of $\tau$ the higher is the value of the ratio $\varphi_e/\tau$, 
therefore a shorter time delay $\Delta t$ is need to fulfill auto-ignition condition (\ref{restart}).
Ignition by contact with the flame occurs after a time of the order of $10 \, \rms$ \cite{fletcher_etal-cst-2007,pickett_etal-ijwf-2010}.
Here, since auto-ignition occurs as a consequence of heat convection,
the characteristic time for auto-ignition is stated to be $2$ and $3$ orders of magnitude longer 
than that for a direct contact with the flame. 
Numerical solutions are obtained when $\tau=600 \, \rms$, $3000 \, \rms$, $6000 \, \rms$.
In the case of the ordinary level-set motion,
$\tau$ follows to be equal to the time after which
a certain point is reached and instantaneously burned by the firefront, see the analysis in Section 3.

ROS $\mV(\obx,t)$ is calculated from the well-known Rothermel semi-empirical formula \cite{rothermel-1972}, 
which is an operative approximation of a theoretically based formula due to Frandsen \cite{frandsen-cf-1971},
\be
\mV(\obx,t)= \mV_0 (1+f_W+f_S) \,,
\ee
where $\mV_0$ is the spread rate in the absence of wind, $f_W$ is the wind factor and $f_S$ is the slope factor.
For the full description of $\mV_0$, $f_W$ and $f_S$, 
the reader is referred to fireLib and Fire Behaviour SDK software documentation ({\tt http://fire.org}) and to Reference \cite{mandel_etal-gmd-2011}.
However, to best highlight the model performance, simulations are carried out in the most simple case 
with no wind, no slope and short grass fuel, i.e. NFFL (Northern Forest Fire Laboratory) Model 1,
and with a unique dead fuel moisture of the type $1$-hour dead fuel moisture 
(i.e. those fuels whose moisture content reaches equilibrium with the surrounding atmosphere within $1$ hour)
that is stated equal to $0.1$ {\rm [Kg water/Kg fuel]}.

The considered mesh size is $h=30.48 \, \rmm$ and the simulation domain is builted up with $100$ cells in each direction.
The time step is computed according to (\ref{dt}) and it follows to be $dt=2.2824 \times 10^3 \, \rms$. 
A constant time step is due to the simple conditions of the case studies and to the fixed grid spacing.
The computational time is of the order of few seconds on a standard personal computer.

\subsection{Pre-heating and its accelerating action on the firefront propagation}
Wildland fire propagation depends on two competing ignition mechanisms that are: 
the arrival of the flame in a certain place and the hot-air heating.
If the auto-ignition time $\Delta t$ is short enough, locations heated by the hot air can burn before than the fire flame is arrived 
and the effective firefront velocity results to be increased, see Section \ref{sec: planefront}.
Hence, the pre-heating accelerates the wildland fire propagation.

\begin{figure}
\includegraphics[width=7cm]{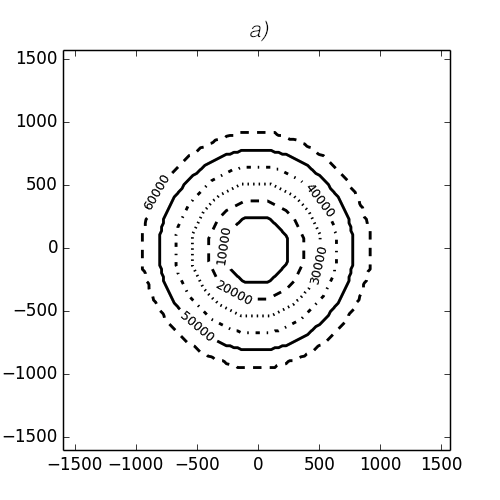}
\includegraphics[width=7cm]{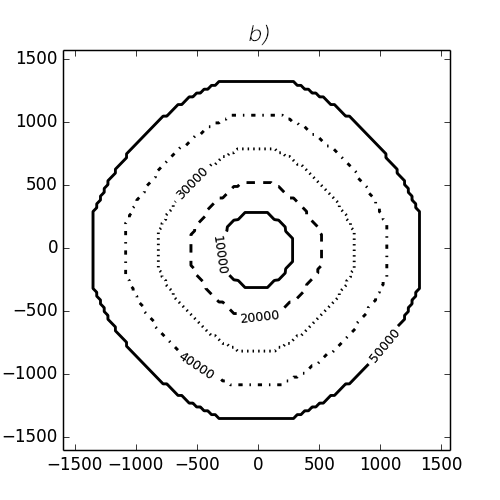} 

\includegraphics[width=7cm]{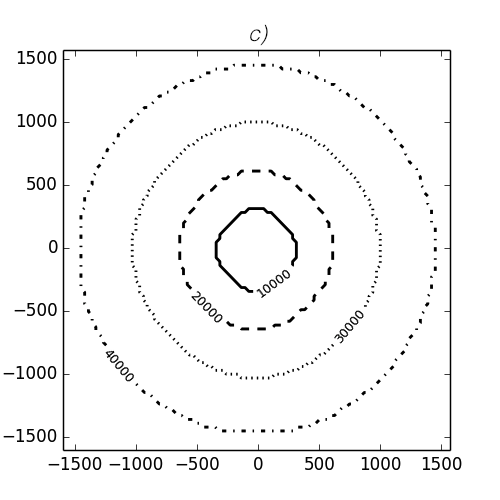}
\includegraphics[width=7cm]{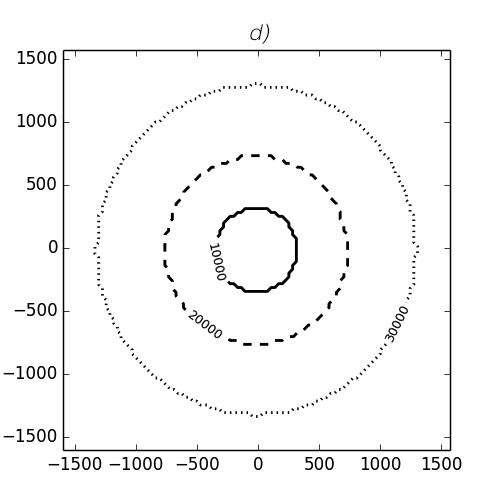}
\caption{Evolution in time of the fireline contour, when $\tau=600 \, \rms$, 
for the level-set method $a)$ and for the randomized level-set method with increasing turbulence: 
$b)$ $\mD=4 \times 10^{-2} \, \rmm^2\rms^{-1}$, $c)$ $\mD=1.5 \times 10^{-1} \, \rmm^2\rms^{-1}$,
$d)$ $\mD=3.5 \times 10^{-1} \, \rmm^2\rms^{-1}$.
The domain axes are expressed in feets and numbers labelling lines refer to the elapsed time in minutes.}
\label{acct10}
\end{figure}

\begin{figure}
\includegraphics[width=7cm]{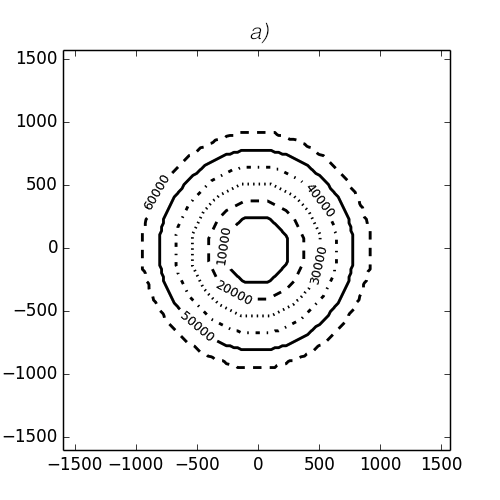}
\includegraphics[width=7cm]{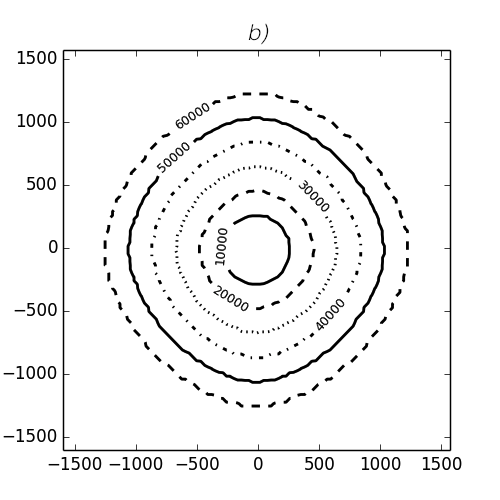}

\includegraphics[width=7cm]{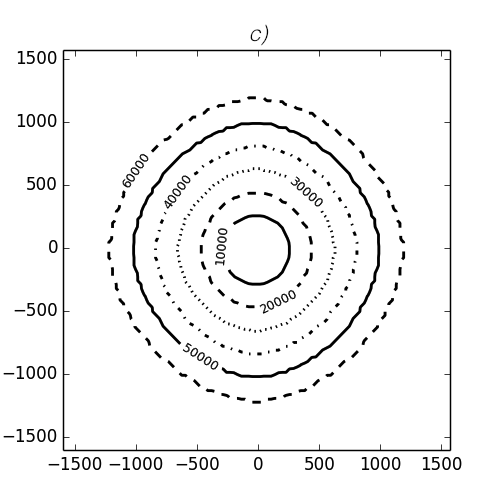}
\includegraphics[width=7cm]{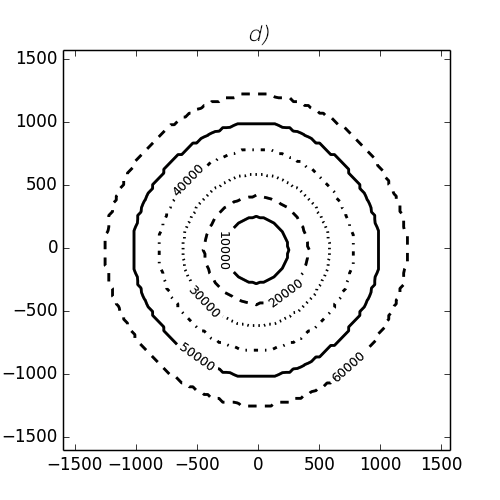}
\caption{The same as in {\it Figure \ref{acct10}} but when $\tau=3000 \, \rms$.}
\label{acct50}
\end{figure}

This acceleration character of the heating-before-ignition is shown in Figures \ref{acct10} and \ref{acct50} 
where strong turbulence and short characteristic time $\tau$ generate a faster firefront propagation, 
whereas for long $\tau$ the effects of the pre-heating are negligible.  
In the plots, the strong or weak effect of pre-heating is embodied by the time required by the fire 
to reach the boundaries of the considered domain.
In fact, in all cases in Figure \ref{acct10} where $\tau=600 \, \rms$, 
it is evident that, with the proposed formulation, the firefront spreads faster than with the ordinary level-set approach. 
Borders of the domain are reached by the fire contour in a temporal interval that decreases by increasing the turbulent diffusion coefficient.
Differently, in Figure \ref{acct50} where $\tau=3000 \, \rms$, 
the effects of the increasing of the turbulent diffusion coefficient are reduced.
But a faster propagation than the ordinary level-set approach is kept. 

\subsection{Fire front overcoming firebreaks}
\begin{figure}[ht]
\includegraphics[width=7cm]{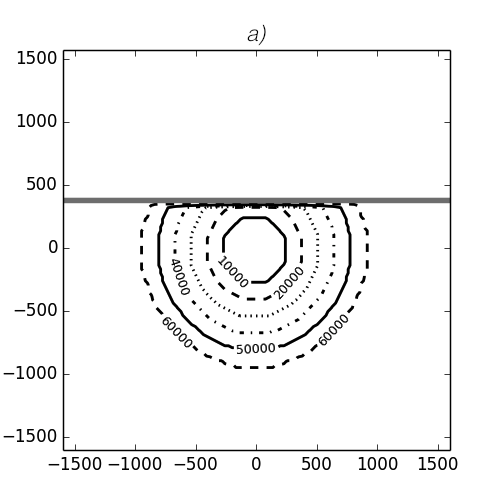}
\includegraphics[width=7cm]{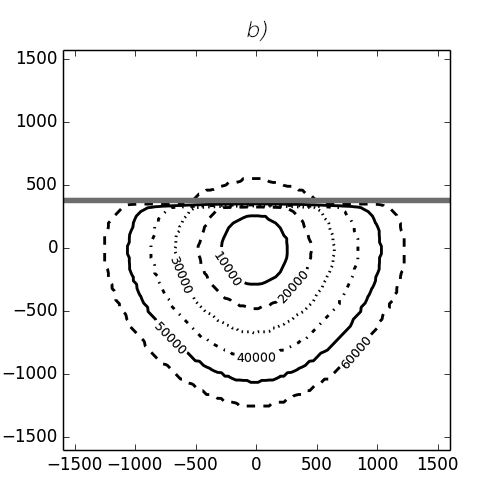}

\includegraphics[width=7cm]{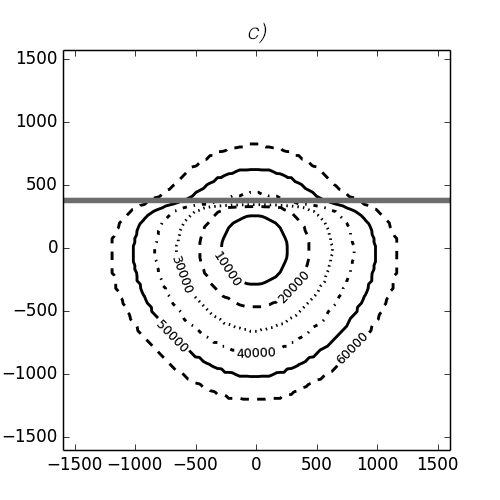}
\includegraphics[width=7cm]{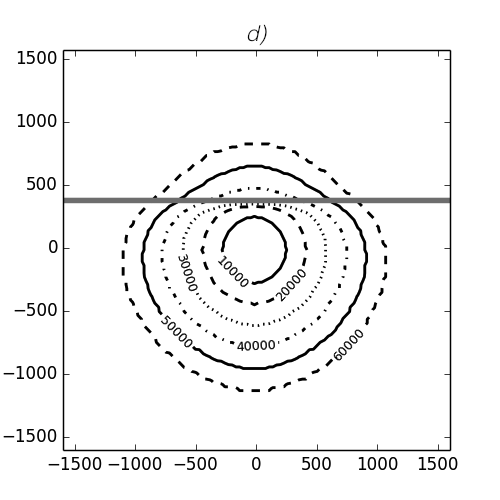}
\caption{The same as in {\it Figure \ref{acct10}} but in the presence of a firebreak and when $\tau=6000 \, \rms$.}
\label{firebreak}
\end{figure}

The ordinary level-set method fails when managing the real situation of a fire that overcomes a firebreak.
In fact, the firebreak is a zone without fuel so that $\mV(\bx,t)=0$, which causes the fire to stop.
However, the hot air mass can overcome the firebreak and ignite an area after it, so that a new firefront starts. 
In Figure \ref{firebreak} it is shown the suitability of the proposed model to simulate the hot air 
that overcomes a firebreak, which is represented by an horizontal strype, and a new fire ignited. 
The stronger is the turbulence the earlier is the ignition behind the firebreak. 

However, it is here reminded that fire can overcome fire breaks also because of fire spotting 
that requires modelling as well, see e.g. \cite{bhutia_etal-james-2010}.
In future, this issue will be considered within the proposed approach to improve the present model.

\section{Conclusion}
\label{sec: conclusion}
Turbulence has an important role in wildland fire propagation because,
together with other phenomena as for example fire spotting, vegetation distribution, gaseous combustion fluctuation and 
small-scale terrain elevation changes, it gives a random motion to the fireline propagation.
Turbulence transports hot air mass generated by the fire, which can pre-heat and ignite the zones ahead the fire.
Then it aids the enlarging of the burned area. 

Modeling turbulence effects in wildland fire propagation has been here considered.
The suitability of an approach coupling the level-set method for tracking fireline contour and the turbulent diffusion of the hot mass air 
has been investigated. This approach is named randomized level-set method. 

Turbulent transport has been inserted into the level-set approach by randomizing the position of the contour points according to 
the PDF of the hot air particles and keeping the ordinary level-set contour as their average position.
The wildland fire propagation follows to be the weighted distribution of the ordinary level-set contour whose weight function is 
the particle PDF.
 
Actually this formulation coincides with the Lagrangian algorithm used to numerically solve the level-set equation 
that adopts a kernel function to smooth the level-set isoline properties. 
In the present approach the choice of the kernel function is removed, since it is the particle PDF and, 
with the assumption of a parabolic model for turbulent diffusion (\ref{parabolic}), the kernel function is the Gaussian PDF (\ref{gaussian2D}).

By comparing the ordinary and the randomized level-set methods,
it is emerged that the growth of the burned area can be faster by the randomized approach 
than by the ordinary approach because of the pre-heating action.
When the pre-heating action is not taken into account, 
the front propagation by the randomized level-set is always slower than that by the classical level-set approach. 
This is a consequence of the diffusion process.

The proposed formulation is appeared to be suitable to manage the following two dangerous situations: 
$i)$ the faster propagation of the fireline as a consequence of the pre-heating action by the hot air mass,
$ii)$ the overcoming of a break-fire by the fire because of the diffusion of the hot air behind it.
In fact, since the firebreak is a zone without ground fuel, 
the ROS follows to be null and this causes the fire stop when the ordinary level-set approach is used.
But, the hot air mass can overcome the firebreak and ignite an area behind it, 
so that a new firefront starts, as reproduced by the randomized approach.
However, it is remarked that fire overcoming a firebreak can accour also because of fire spotting, see e.g. \cite{bhutia_etal-james-2010}. 

The improvement of the discussed model can be done by including further effects 
that give a random character to the firefront propagation 
(e.g. spotting phenomena, vegetation distribution, gaseous combustion fluctuation, small-scale terrain elevation changes).
This constitutes the future development of the research.

\section*{Acknowledgments}
Authors would like to thank 
colleagues Ernesto Bonomi, Eva Lorrai, Marino Marrocu and Antioco Vargiu at CRS4 
are also acknowledged for fruitful discussions 
and Giuseppe Delogu of ``Corpo forestale e di vigilanza ambientale della Regione Sardegna'' for the precious guidelines.
GP research is presently supported by the Basque Government through the BERC 2014--2017 program and by 
the Spanish Ministry of Economy and Competitiveness MINECO: BCAM Severo Ochoa accreditation SEV--2013--0323.

\bibliographystyle{plain} 
\bibliography{wildlandfire}
\end{document}